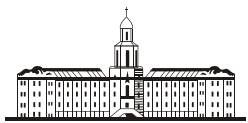
| РОССИЙСКАЯ АКАДЕМИЯ НАУК | RUSSIAN ACADEMY OF SCIENCES |
|---|---|
| **ИНСТИТУТ ПРОБЛЕМ БЕЗОПАСНОГО РАЗВИТИЯ АТОМНОЙ ЭНЕРГЕТИКИ** | **NUCLEAR SAFETY INSTITUTE** |

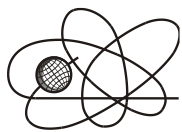

Препринт ИБРАЭ № IBRAE-2018-12           Preprint IBRAE-2018-12

**Arutyunyan R.V., Osadchy A.V.**

# THE SYSTEMS OF VOLUME-LOCALIZED ELECTRON QUANTUM LEVELS OF CHARGED FULLERENES

Москва                                   Moscow
2018                                     2018

# The systems of volume-localized electron quantum levels of charged fullerenes

Rafael V. Arutyunyan[1], Alexander V. Osadchy[1]

[1] Nuclear Safety Institute of the Russian Academy of Sciences (IBRAE RAN), 52, Bolshaya Tulskaya Street, Moscow, 115191, Russia

Correspondence should be addressed to Alexander V. Osadchy; aosadchy@kapella.gpi.ru

## Abstract

The existence of a system of short-live discrete volume-localized electron quantum levels in positively charged fullerenes is theoretically and numerically demonstrated using the example of fullerenes $C_{60}$ and $C_{20}$. Unlike well-studied experimentally and theoretically electron states localized in a thin surface layer, these electron states are due to the flat part of the Coulomb potential of a positively charged fullerene sphere. The energy width of the system of such discrete volume-localized levels depends on the charge and increases with charge increasing. For $C_{60}^{+1}$, the energy width is 0.32 a.u. and increases up to 1.9 a.u. for fullerene $C_{60}^{+10}$. Thus, the electrons captured on these discrete levels of fullerene form a sort of a short-lived "nano-atom" or "nano-ion", in which the electrons are localized inside a positively charged spherical "nucleus". Numerous published papers have demonstrated theoretically and experimentally the existence of metastable positively charged $C_{60}$ fullerenes with a charge of +10 or more, which suggests the possibility of experimental observation of the considering system of volume-localized electronic states.

## Introduction

Fullerenes are well experimentally and theoretically investigated object. A large number of papers have been devoted to the study of the electronic states of neutral fullerenes [1-4]. Various methods have been used to study the stability and the mechanism of the decay of charged fullerenes. A number of experimental studies demonstrate the existence of a metastable $C_{60}^{+n}$ cation with charges up to +10 and more [5-10]. These results were observed in the collision of fullerenes with highly charged ions. Also, it was shown in the works that charged $C_{60}$ molecules are most likely to decay by emission of $C^{+2}$ [5-7] and with a lower probability of $C^{+4}$ [6,7]. The largest value of the $C_{60}$ charge, observed experimentally, is +12 [8]. This result was achieved by irradiating $C_{60}$ molecules with intense laser radiation.

A theoretical study of the stability limit of fullerene molecules is presented in papers using different approaches [5-7, 9]. In the work based on the Dirac-Fock-Slater simulation [9], the limiting charge is +13. The application of the molecular dynamics method, together with a simplified approach based on the density functional theory, demonstrates the limiting charge from +16 to +19 [10]. Calculations of the lifetime of highly charged (+10 or more) fullerenes are in the range of microseconds [11] to seconds [12].

In this paper, the existence of a system of short-live discrete volume-localized electron quantum levels in positively charged fullerenes is shown theoretically and numerically in the example of fullerenes $C_{60}$ and $C_{20}$ on the basis of the theoretical approach presented in [13].

**Methods and approaches**

The modeling was carried out using two main approaches. As the first, a numerical solution of the Schrodinger equation for a spherically symmetric potential in the nodal approximation was applied. Most calculations were carried out using the calculation zone 50 a.u. and the number of nodes is 1000.

The second approach was based on the density functional theory (DFT) [14], which was implemented in the software package QuantumEspresso [15]. The electron wave functions are decomposed in a plane wave basis. To reduce the dimension of the plane wave basis, the pseudopotential method was used. In the study of nano-sized materials, the supercell method with a translation vector length of 100 au was used to exclude the interaction between fullerenes. As the pseudopotential, the Perdew-Wang norm concerving potentials [16] were used in the framework of the local density approximation (LDA). The basis takes into account plane waves with energies less than 40 Ry. The structure was optimized using a method based on the Broyden-Fletcher-Goldfarb-Shanno algorithm. The positions of the ions varied to a state where the interatomic forces became less than $10^{-4}$ Ry/au, and the parameters of the unit cell varied to values at which the stress in the cell became less than 0.5 Kbar. Calculations were carried out on a high-performance cluster computer K-100 of the M.V. Keldysh Mathematics Institute of Russian Academy of Sciences.

**Results and discussion**

The simplest physical model for describing the potential of a charged fullerene is the widely used approximation of the charged sphere field potential:

$$U(r) = -Z\, U_\Phi(r), \quad U_\Phi(r) = \begin{cases} \dfrac{1}{R_f} & \text{when } r \leq R_f \\ \dfrac{1}{r} & \text{when } r \geq R_f \end{cases}, \tag{1}$$

where Z – positive charge, and $R_f$ – fullerene radius.

For convenience, we used a dimensionless system of units, assuming the electron mass m = 1, the electron charge e = 1, ℏ = 1. In general, our attention will be directed to the study of the fullerene $C_{60}$, whose radius we take $R_f = 6.627$ a.u. [1].

A simple estimate of the discrete energy levels of an electron in such a potential can be obtained from the known solutions of the Schrödinger equation for a spherical rectangular well of depth $U_0 = \dfrac{Z}{R_f}$. Within this sphere ($0 \leq r \leq R_f$), the solution of the Schrödinger equation is described by a spherical Bessel function $\chi = j_l\left(\dfrac{\xi r}{R_f}\right)$, that satisfies the boundary conditions at zero $\chi(0) = 0$. Then, as outside the well ($R_f < r < \infty$), the solution that satisfies $\chi(\infty) = 0$, is represented by the spherical Henkel function $\chi = h_l\left(\dfrac{i\eta r}{R_f}\right)$. The parameters ξ and η are algebraically related:

$$\xi^2 + \eta^2 = 2U_0 R_f^2 \tag{2}$$

and determine the discrete energy levels:

$$E_n = -\dfrac{\eta^2}{2R_f^2} = -U_0 + \dfrac{\xi^2}{2R_f^2}, \tag{3}$$

The values of the parameters ξ and η are fixed by the condition of continuity of the wave function for $r = R_f$. For l=0 it is means:

$$\eta = -\xi\, \text{ctg}\, \xi$$

The tables 1 and 2 show the electron levels of the spherical potential well at Z=1 and Z=5, obtained from (3). Also in the tables, for comparison, the results of numerical solutions of the Schrödinger equation for a very deep well $U_0 \gg \dfrac{1}{2R_f^2}$ and a well of finite depth are shown.

To calculate the energy spectrum of electrons in the potential well (1), we solved the standard Schrödinger equation for the radial component of the wave function:

$$\dfrac{d^2\chi}{dr^2} - \dfrac{l(l+1)}{r^2}\chi + 2(E - U(r))\chi = 0 \text{ , where} \tag{4}$$

$$\chi(r) = rR(r)$$

The calculations has been performed in the nodal approximation with the number of nodes n = 1000.

| l | Analytical | Numeric | |
|---|---|---|---|
| | | **Very deep well** | **Finity depth** |
| | **Energy, a.u.** | **Energy, a.u.** | **Energy, a.u.** |
| 0 | -0.039 | -0.03951 | -0.08491 |
| 1 |        |          | -0.02238 |

Table 1. Energy levels of a rectangular spherically symmetric well for Z = 1.

| l | Analytical | Numeric | |
|---|---|---|---|
| | | **Very deep well** | **Finity depth** |
| | **Energy, a.u. (analytical)** | **Energy, a.u. (numeric)** | **Energy, a.u.** |
| 0 | -0.642 | -0.64259 | -0.66606 |
| 1 | -0.525 | -0.52419 | -0.57418 |
| 2 | -0.376 | -0.37602 | -0.45974 |
| 0 | -0.305 | -0.30886 | -0.40861 |
| 3 | -0.199 | -0.19842 | -0.32414 |
| 1 | -0.075 | -0.07506 | -0.23738 |
| 4 |        |          | -0.23738 |
| 2 |        |          | -0.05561 |
| 0 |        |          | -0.02611 |

Table 2. Energy levels of a rectangular spherically symmetric well for Z = 5.

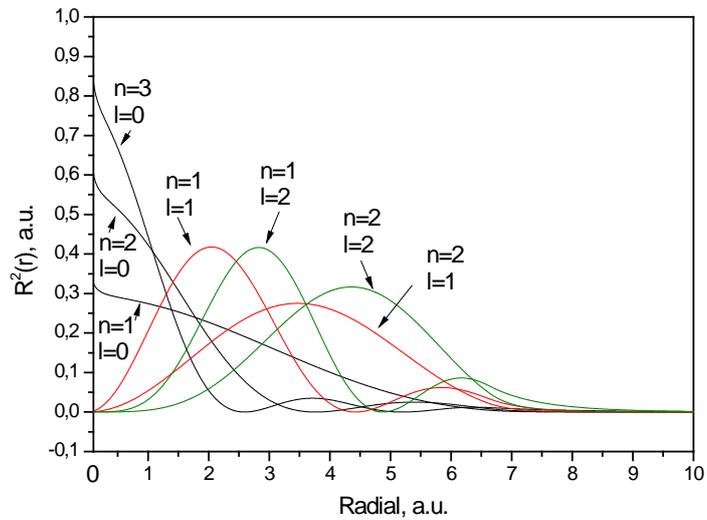

Figure 1. Wave function radial component squares for electron states of a rectangular spherically symmetric well of finite depth at Z = 5 received as a result of the numerical calculations of the Schrödinger equation (4)

Comparing the results shown in Tables 1 and 2 it can be seen that the discrete energy levels obtained by an analytical solution of the Schrödinger equation for an infinitely deep spherically symmetric well and by numerical solution in the nodal approximation coincide with good accuracy. Numerical solutions for a well of finite and infinite depth coincide for the lowest-lying energy levels and expectly begin to diverge with increasing energy.

A characteristic feature of the system of discrete levels corresponding to the smooth part of the potential (2) is that the electron wave functions corresponding to them are localized in the fullerene volume, in contrast to the well-studied surface-localized electronic states (Fig. 1). The number of such states increases with increasing depth of the potential, which occurs with the growth of the fullerene charge Z (Tables 1 and 2).

To solve the stationary Schrödinger equation (4) for an electron in the potential well (1), we can use a numerical solution for a spherically symmetric potential in the nodal approximation. Figure 2 shows the numerical calculation results obtained for the charge Z = 5 for the number of nodes n = 1000 and the calculation region r≤50 a.u.

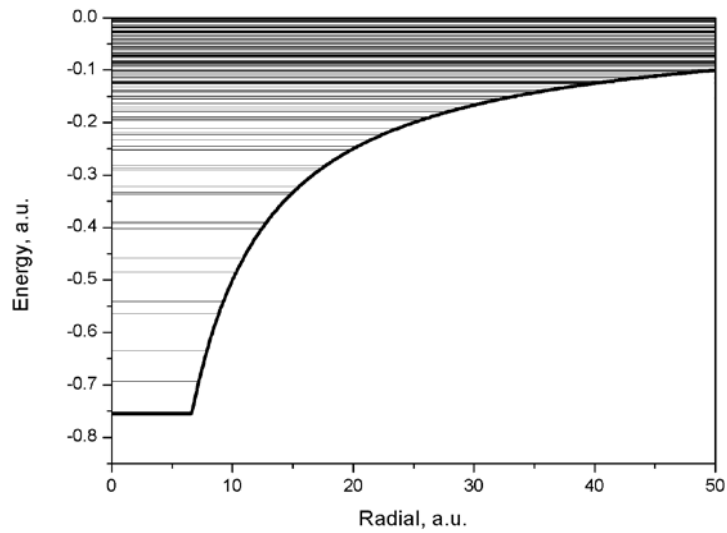

Figure 2. Energy levels of a spherically symmetric potential (1) at Z = 5 received as a result of the numerical calculations of the Schrödinger equation (4)

## Potential of the charged fullerene taking into account the Coulomb field and the analytic approximation of the well on the surface in the model of the jelly

For the subsequent consideration of the discrete electronic levels of a charged fullerene, we take into account the potential as the sum of the Lorentz potential of the surface layer in the jelly model [17] and the Coulomb potential of the positively charged sphere.

$$U(r) = -\frac{V}{(r-R)^2+d^2} - ZU_\Phi(r) \qquad (5)$$

, where, in accordance with [17] V=0.711, R=6.627, d=0.610 and Z – positive charge of fullerene

In this case, we neglect the influence of the Coulomb potential (1) on the potential (5).

The potentials (5) for different Z are shown in Fig. 3. The potential in the center of the fullerene becomes deeper as the charge increases. For Z = 1 U (0) = - 0.17 au, whereas in the case of Z = 5 U (0) = - 0.77 au. and U (0) = - 1.52 a.u. at Z = 10. As a result, we obtain two types of electronic states: localized on a thin sphere of fullerene and with a volumetric localization of the electron.

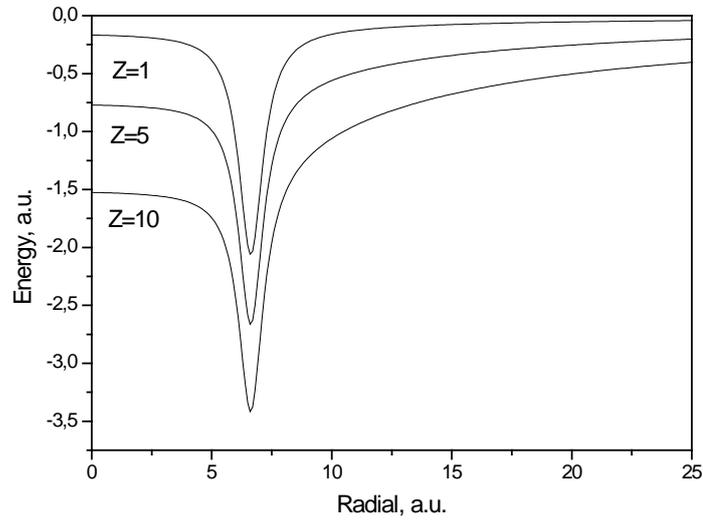

Figure 3. Analytical potential of fullerene (5) for various Z.

This is clearly seen in the results of numerical solutions of the Schrödinger equation for an electron in the potential (5) for different Z. Figure 4 shows the spectra of a charged fullerene at Z = 1 and Z = 5. Tables 3 and 4 contain a set of discrete electron states for a charged fullerene at Z = 1 and Z = 5, respectively. For comparison, tables 3 and 4 show discrete electronic states for the Coulomb potential of the charged sphere (1) for Z = 1 and Z = 5, respectively.

| Potential (1) | | | Potential (5) | | |
|---|---|---|---|---|---|
| n | l | Energy, a.u. | n | l | Energy, a.u. |
|  |  |  | 1 | 0 | -1.05864 |
|  |  |  | 1 | 1 | -1.01158 |
|  |  |  | 1 | 2 | -0.94142 |
|  |  |  | 1 | 3 | -0.84862 |
|  |  |  | 1 | 4 | -0.73372 |
|  |  |  | 1 | 5 | -0.59734 |
|  |  |  | 1 | 6 | -0.44019 |
|  |  |  | 1 | 7 | -0.26313 |
|  |  |  | 1 | 8 | -0.18781 |
|  |  |  | 2 | 0 | -0.14966 |
| 1 | 0 | -0.11313 | 2 | 2 | -0.10578 |
| 1 | 1 | -0.08074 | 2 | 0 | -0.07869 |
| 2 | 0 | -0.05527 | 3 | 9 | -0.06724 |
| 1 | 2 | -0.05132 | 1 | 3 | -0.06289 |
| 2 | 1 | -0.04079 | 2 | 1 | -0.05440 |
| 3 | 0 | -0.03175 | 3 | 0 | -0.03975 |
| 1 | 3 | -0.03104 | 4 | 2 | -0.03907 |
| 2 | 2 | -0.02903 | 3 | 1 | -0.02987 |
| 3 | 1 | -0.02427 | 4 | 4 | -0.02805 |
| 1 | 4 | -0.01974 | 2 | 3 | -0.02762 |

Table 3. Energy levels of a spherically symmetric potential (1) and potential (5) for Z = 1 received as a result of the numerical calculations of the Schrödinger equation (4)

| Potential (1) | | | Potential (5) | | |
|---|---|---|---|---|---|
| n | l | Energy, a.u. | n | l | Energy, a.u. |
| | | | 1 | 0 | -1.66521 |
| | | | 1 | 1 | -1.64135 |
| | | | 1 | 2 | -1.59381 |
| | | | 1 | 3 | -1.52295 |
| | | | 1 | 4 | -1.42919 |
| | | | 1 | 5 | -1.31308 |
| | | | 1 | 6 | -1.17518 |
| | | | 1 | 7 | -1.01614 |
| | | | 1 | 8 | -0.83671 |
| | | | 2 | 0 | -0.75648 |
| 1 | 0 | -0.69530 | 2 | 1 | -0.69254 |
| 1 | 1 | -0.63561 | 1 | 9 | -0.63773 |
| 1 | 2 | -0.56490 | 2 | 2 | -0.61415 |
| 2 | 0 | -0.54226 | 3 | 0 | -0.55998 |
| 1 | 3 | -0.48600 | 2 | 3 | -0.53094 |
| 2 | 1 | -0.45972 | 3 | 1 | -0.48054 |
| 1 | 4 | -0.40322 | 2 | 4 | -0.44853 |
| 3 | 0 | -0.39317 | 1 | 10 | -0.42032 |
| 2 | 2 | -0.39088 | 3 | 2 | -0.40964 |
| 3 | 1 | -0.33771 | 4 | 0 | -0.39563 |

Table 4. Energy levels of a spherically symmetric potential (1) and potential (5) for Z = 5 received as a result of the numerical calculations of the Schrödinger equation (4)

It can be seen from the tables that the numerical values of the energy of the discrete excited levels of the Coulomb potential (1) and the values of the levels of the total potential (5) located above the plane part of the Coulomb potential are fairly close.

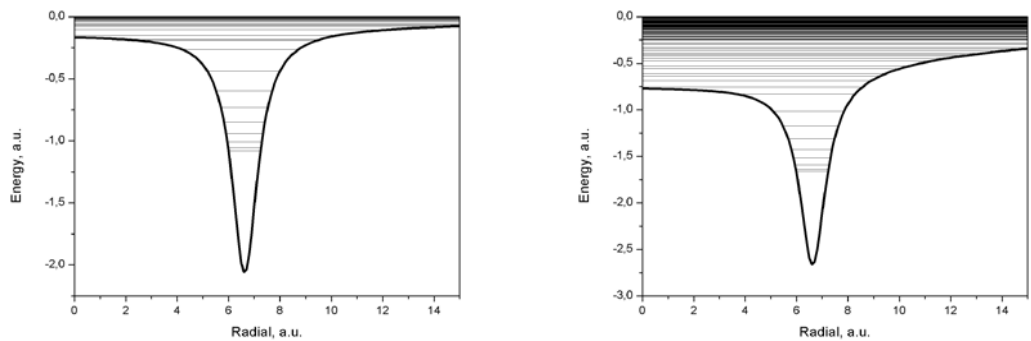

Figure 4. Energy levels of a spherically symmetric potential (5) at Z = 1 (left) and Z = 5 (right) received as a result of the numerical calculations of the Schrödinger equation (4)

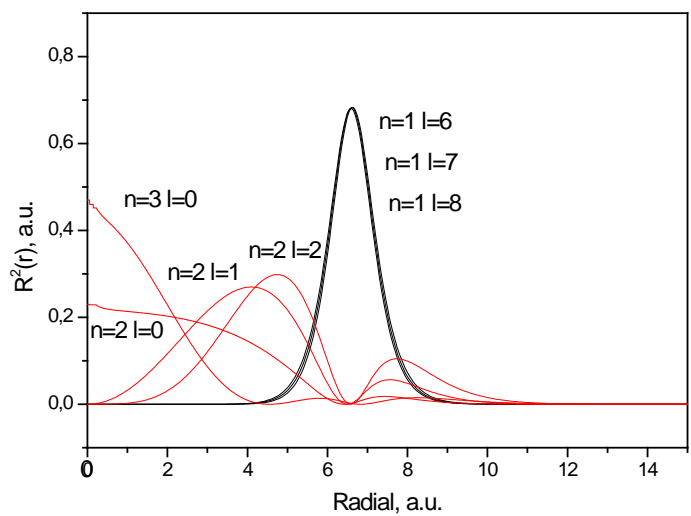

Figure 5. Squares of the wave functions radial component of the spherically symmetric potential (5) states at Z = 5 received as a result of the numerical calculations of the Schrödinger equation (4)

Figure 5 shows the results of numerical calculations of the squares of the radial components of the wave functions of a spherically symmetric potential (5) at Z = 5. As can be seen from the graphs, along with well-known surface-localized levels in the energy spectrum of electrons, there are volume-localized excited levels.

# The calculation of energy levels of discrete states of a charged $C_{60}$ fullerene on the basis of the electron density functional

To compare the results obtained, numerical calculations of the potentials of charged fullerenes were carried out using a method based on the electron density functional theory. Numerical three-dimensional potentials for the electron in a charged fullerene are obtained. In contrast to potential (5), these dependencies take into account the positions of each carbon atom, which leads to a violation of the spherical symmetry. It is necessary to distinguish two characteristic cross sections of the potentials obtained: passing through the center of the fullerene and through the carbon atom, and passing through the middle of the segment connecting neighboring atoms. Figure 6 contains the analytical potentials for the $C_{60}$ fullerene at Z=3 and Z=5, as well as the cross sections calculated numerically by the DFT potential, passing through the carbon atom and through the center of a segment between neighboring atoms.

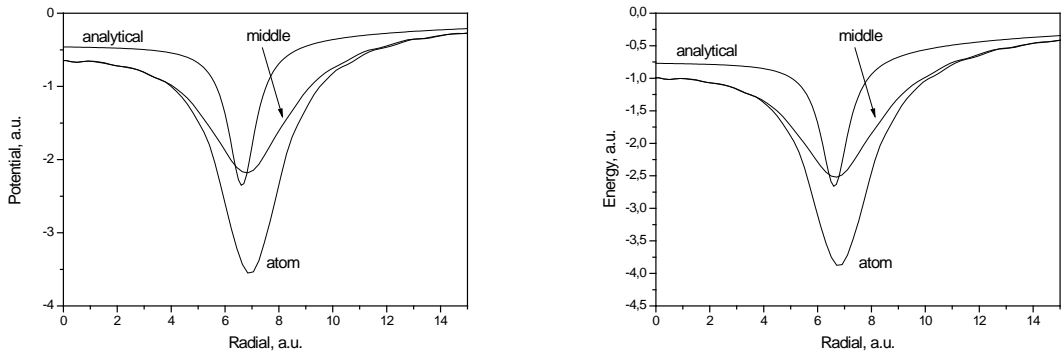

Figure 6. Comparison of the analytical potential of the charged fullerene (5) at Z = 3 (on the left) and Z = 5 (on the right) with cross sections of the calculated potential at Z = 3 and Z = 5, passing through the atom of carbon (*atom*) and through the center of the segment connecting the neighboring atoms (*middle*).

It can be seen that the maximum depth of the potential well obtained using DFT, similar to the potential (5), decreases with increasing Z. The value of the potential in the center of the fullerene is smaller than the analytical approximation (5) that we have used. The potential well located on the radius of the fullerene is much deeper and wider. The reason for this difference is the relatively rough approximation within the jelly model, whereas the DFT makes it possible to more accurately calculate the potential, taking into account the positions of all fullerene atoms.

We consider the solution of the Schrödinger equation in a centrally symmetric field, where potentials calculated using the DFT method was used as the field potential. The cross sections for the potentials under consideration were taken through the carbon atoms and passing through center of a segment between the neighboring atoms. Despite the fact that the real field of fullerenes is not centrally symmetric, this approach may be justified because we are most interested in states with energies above the potential well in the center of the charged fullerene. These states are the least subject to the lack of symmetry.

Figure 7 shows the examples of the obtained electronic states for $C_{60}$ fullerene with charges +1, +5 and +10.

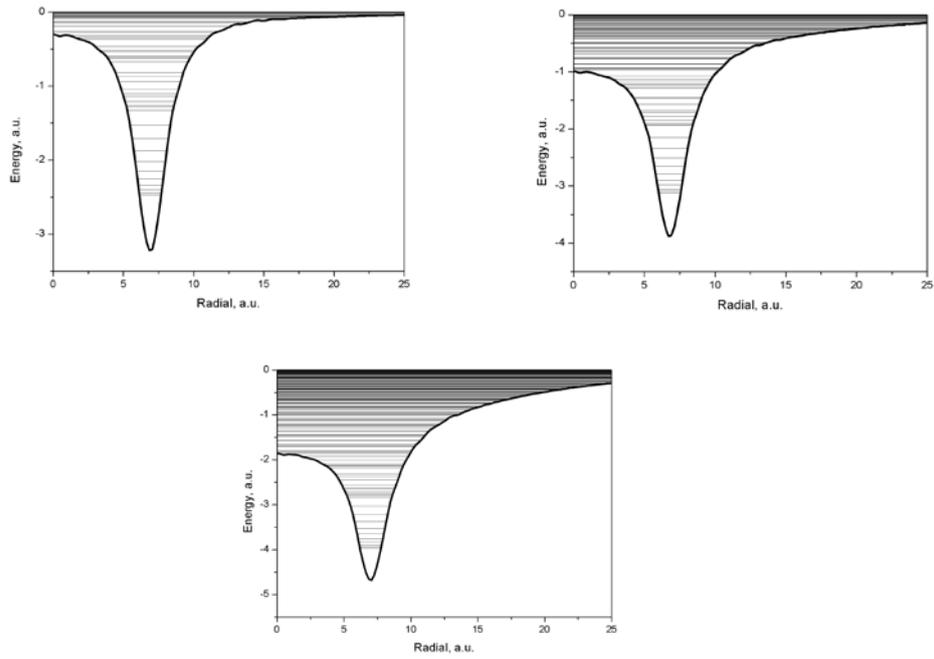

Figure 7. Energy levels in spherical symmetric potential received as a result of the numerical calculations of the Schrödinger equation (4) using the DFT method for the $C_{60}$ fullerene with charge +1 (top left), +5 (top right) and +10 (bottom)

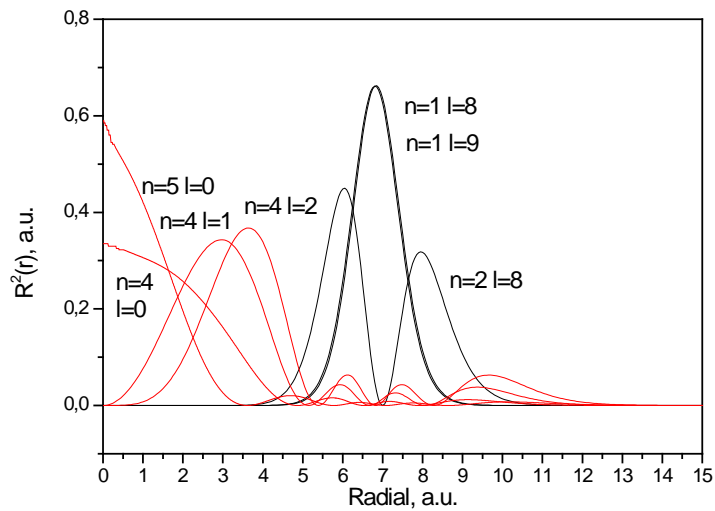

Figure 8. Squares of the radial component of the wave functions of states of a spherically symmetric potential received as a result of the numerical calculations of the Schrödinger equation (4) using the DFT method for $C_{60}$ fullerene at $Z = 5$

The figure 8 presents the results of numerical calculations of the squares of the radial components of the wave functions of a spherically symmetric potential calculated for a C60 fullerene with a charge of +5. Analogously to the case of the analytical potential (5), it can be seen from the graphs that along with the well-known surface-localized levels in the energy spectrum of the electrons, there are volume-localized excited levels.

As can be seen from figures 7 and 8, the energy spectra of the electrons and wave functions calculated using the potential obtained by the method based on the electron density theory are close to the results obtained in solving the Schrödinger equation for analytical potentials (5). The principal difference between the results obtained for the potential (5) and the calculated one using DFT, is observed for states located below the potential in the center of the fullerene. These states are localized within the thin sphere of the fullerene surface. Due to the fact that DFT takes into account the positions of all fullerene atoms in an explicit form, the potential are not spherically symmetric, as in the case with the potential (5). At the same time, states with energies greater than the potential at the center of the fullerene are localized in the fullerene volume, and the calculated potential for the DFT is practically spherically symmetric, analogous to (5). This is clearly seen in figure 8, which represents the calculated squares of the wave functions obtained as a result of the solution of the Schrödinger equation for states located at the level of the boundaries of occupied and free states at the bottom of the potential well at the center of the fullerene and above the bottom of the potential well. Data for fullerene $C_{60}$ with charge +10 are presented. Fullerene with other charges demonstrates similar results.

## Discrete energy levels of charged $C_{20}$ fullerene

Similar results can be obtained for other types of fullerenes, including those for $C_{20}$. According to [18], the radius of a given fullerene can be taken equal to R = 2.93 a.u. The electron states obtained by numerically solving the Schrödinger equation for a spherically symmetric potential (1), applied to $C_{20}$, are shown in figure 9, the values of 10 states with the lowest energy are shown in table 5.

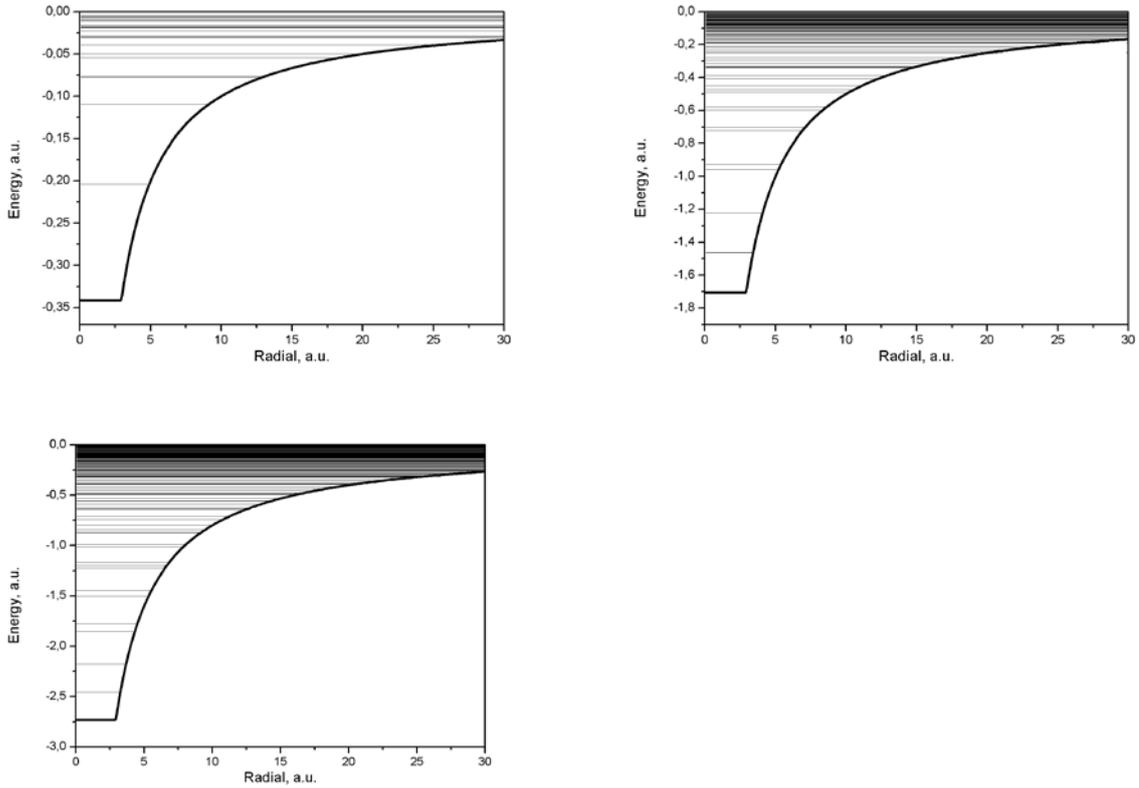

Figure 9. Energy levels of a spherically symmetric potential (1) received as a result of the numerical calculations of the Schrödinger equation (4) for $C_{20}$ fullerene at $Z = 1$ (top left), $Z = 5$ (top right) and $Z = 8$ (bottom).

| Z=1 | | | Z=5 | | | Z=8 | | |
|---|---|---|---|---|---|---|---|---|
| n | l | Energy, a.u. | n | l | Energy, a.u. | n | l | Energy, a.u. |
| 1 | 0 | -0.2042 | 1 | 0 | -1.4612 | 1 | 0 | -2.4539 |
| 1 | 1 | -0.1104 | 1 | 1 | -1.2205 | 1 | 1 | -2.1725 |
| 2 | 0 | -0.0776 | 1 | 2 | -0.9592 | 1 | 2 | -1.8511 |
| 1 | 2 | -0.0552 | 2 | 0 | -0.9271 | 2 | 0 | -1.7761 |
| 2 | 1 | -0.0506 | 2 | 1 | -0.7247 | 1 | 3 | -1.5060 |
| 3 | 0 | -0.0401 | 1 | 3 | -0.7038 | 2 | 1 | -1.4462 |
| 1 | 3 | -0.0311 | 3 | 0 | -0.6009 | 3 | 0 | -1.2280 |
| 2 | 2 | -0.0310 | 2 | 2 | -0.5781 | 2 | 2 | -1.1997 |
| 3 | 1 | -0.0290 | 1 | 4 | -0.4919 | 1 | 4 | -1.1673 |
| 4 | 0 | -0.0238 | 3 | 1 | -0.4780 | 3 | 1 | -1.0172 |

Table 5. Energy levels of a spherically symmetric potential (1) at $Z = 1$, 5 and 8 received as a result of the numerical calculations of the Schrödinger equation (4)

Calculation by a method based on the theory of the electron density functional theory makes it possible to obtain potentials similar to the case of a charged $C_{60}$ fullerene. The electron states in the spherical symmetric potential calculated by the DFT method for $C_{20}$ fullerene with charge $Z = 5$ and $Z = 8$ are shown in figure 8. It is worth noting that the radius of $C_{20}$ increases with increasing of charge. Thus, according to [18], the radius of neutral fullerene is 2.93 a.u., at the same time, according to calculations using DFT, shown in Fig. 10, with charge Z = 5 it increases to 3.8 a.u., and at Z = 8 up to 4.14 a.u. Figure 11 presents the results of numerical calculations of the squares of the radial components of the wave functions of a spherically symmetric potential calculated for a $C_{20}$ fullerene with a charge of +5. At the same time, we do not consider the question of the stability and lifetime of such positively charged fullerenes.

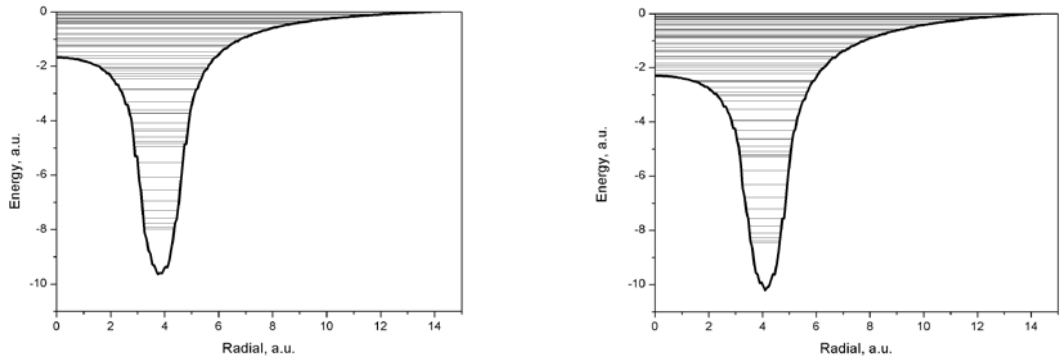

Figure 10. Energy levels in spherically symmetric potential received as a result of the numerical calculations of the Schrödinger equation (4) using the DFT method for C20 fullerene with a charge of +5 (on the left) and +8 (on the right)

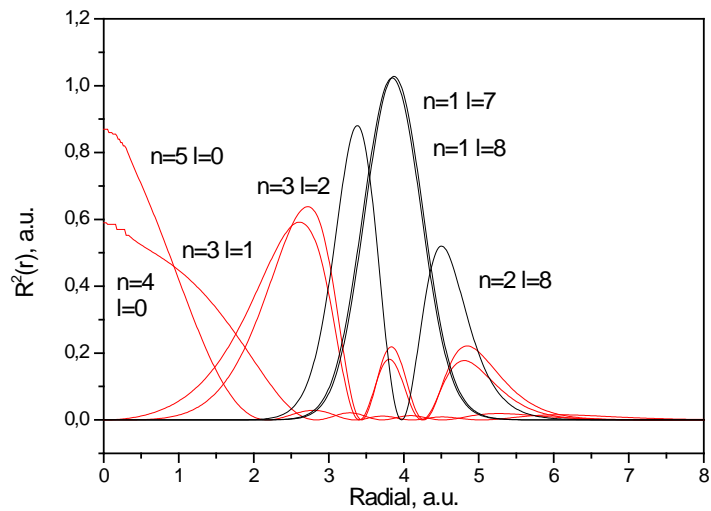

Figure 11. Squares of the radial component of the wave functions of the states of a spherically symmetric potential received as a result of the numerical calculations of the Schrödinger equation (4) using the DFT method for C20 fullerene at Z = 5.

Thus, similarly to $C_{60}$ fullerene, the calculations given for the analytical potential (1) applied to $C_{20}$, as well as numerical solutions of the Schrödinger equation for spherically symmetric potentials of charged $C_{20}$ fullerenes obtained by the DFT method, along with well-studied electron states localized within the thin sphere fullerene, there are states localized in the volume.

## Conclusion

In this paper, we demonstrated on the basis of theoretical and numerical calculations, using the example of fullerenes C60 and C20, the existence of a system of short-lived discrete volume-localized quantum levels of electrons in positively charged fullerenes.

The results obtained provide a consistent qualitative picture for charged fullerenes. To clarify our very approximate results, of course, further research is needed on the basis of microscopic calculations, as well as experimental measurements.

To estimate the lifetime of such levels, one can use the well-known formula for the photon emission rate:

$$P_{ij} = \frac{\omega^3}{3\pi\varepsilon_0 \hbar c^3}|d_{ij}|^2,$$

where $\omega$ – the emission frequency, and $d_{ij}$ – is the matrix element of dipole transition from initial (j) to final (i) state.

One can estimate the magnitude of $d_{ij}$ for transitions between volume-localized levels of charged fullerenes as eR. A simple numerical estimate, taking R=6.627 a.u. for fullerene C$_{60}$ leads to:

$$P_{ij} = \frac{4(6.627)^2}{3\alpha}\frac{c}{\lambda}\left(\frac{\Delta E}{m_e C^2}\right)^3$$

Thus, it is possible to obtain for various $\Delta E$: $P_{ij}(\Delta E = 1\ eV) = 4.67 \cdot 10^7\ s^{-1}$, $P_{ij}(\Delta E = 10\ eV) = 4.67 \cdot 10^{10}\ s^{-1}$. This gives an estimate of the lifetime of states in the range from 21 ns to 21 ps, which is much smaller than the estimated lifetimes of charged fullerenes and suggests the possibility of experimental confirmation of the existence of volume-localized discrete levels.

Experimental confirmation of the existence of these volume-localized discrete levels would be of great interest for experimental research and practical tasks, including the development of new sources of coherent radiation in a wide range of wavelengths.

To estimate the inverse population value n$_{ij}$, which is necessary to reach the threshold for generation of coherent radiation at volume-localized levels of fullerenes, we use the following simple expression:

$$\mu_\omega L_{abs} \gg 1$$

, where

$$\mu_\omega = \frac{\lambda^2}{2\pi}n_{ij}\frac{\Delta\omega}{\Delta\omega_{sp}}$$

$\mu_\omega$ is the resonance amplification factor per unit length;
$L_{abs}$ is the photon loss length;

Δω is full broadening of the emission line due to the Doppler effect, collisional broadening and broadening due to nonradiative losses;

Δω$_{sp}$ is the width of dipole spontaneous emission line;

$\lambda$ transition wavelength between i and j levels; and

ω is the emission frequency at the transition i-j.

$$n_{ij} \gg \frac{2\pi}{\lambda^2 L_{abs}} \frac{\Delta\omega_{sp}}{\Delta\omega}$$

$$\Delta\omega \approx \Delta\omega_{dop} + \Delta\omega_{col} + \Delta\omega_{nrad}$$

The value $L_{abs} \sim \frac{1}{n\sigma_{abs}}$, where n is the volume density of fullerenes. According to experimentally measured values in fullerene pairs of $C_{60}$ [19].

$\sigma_{abs} \sim 10^{-15}\ cm^2$ in the wavelength range of 200-400 nm.

To reach the generation threshold of coherent radiation in the optical range, the estimate of the volume density of the inverted population n$_{ij}$ at the fullerene density n=$10^{14}$-$10^{15}$ cm$^{-3}$ gives the value of $10^{12}$ - $10^{13}$ cm$^{-3}$.